# Experimental evidence for Glycolaldehyde and Ethylene Glycol formation by surface hydrogenation of CO molecules under dense molecular cloud conditions

G. Fedoseev[1, a)], H. M. Cuppen[2], S. Ioppolo[2], T. Lamberts[1, 2] and H. Linnartz[1]

[1]*Sackler Laboratory for Astrophysics, Leiden Observatory, University of Leiden, PO Box 9513, NL 2300 RA Leiden, The Netherlands*

[2]*Institute for Molecules and Materials, Radboud University Nijmegen, PO Box 9010, NL 6500 GL Nijmegen, The Netherlands*

**Abstract**

This study focuses on the formation of two molecules of astrobiological importance - glycolaldehyde (HC(O)CH$_2$OH) and ethylene glycol (H$_2$C(OH)CH$_2$OH) - by surface hydrogenation of CO molecules. Our experiments aim at simulating the CO freeze-out stage in interstellar dark cloud regions, well before thermal and energetic processing become dominant. It is shown that along with the formation of H$_2$CO and CH$_3$OH – two well established products of CO hydrogenation – also molecules with more than one carbon atom form. The key step in this process is believed to be the recombination of two HCO radicals followed by the formation of a C-C bond. The experimentally established reaction pathways are implemented into a continuous-time random-walk Monte Carlo model, previously used to model the formation of CH$_3$OH on astrochemical time-scales, to study their impact on the solid-state abundances in dense interstellar clouds of glycolaldehyde and ethylene glycol.

**Key words:** astrochemistry – methods: laboratory – ISM: atoms – ISM: molecules – infrared: ISM.

## 1 Introduction

Among approximately 180 molecules identified in the inter- and circumstellar medium over 50 molecules comprise of six or more atoms. For astrochemical standards, these are seen as 'complex' species. Most of these molecules contain H-, C- and O- atoms and can be considered as organic molecules. The sources where these complex organic molecules (COMs) are detected include cold interstellar cores, circumstellar envelopes around evolved stars, hot cores and corinos, outflows as well as other regions. The detection of COMs around young stellar objects (YSOs) in the early stages of their evolution may indicate that COMs are part of the material from which comets, planetesimals and ultimately planets are made. Therefore, it is not surprising that likely important prebiotic molecules and chemical pathways leading to their formation at prestellar stages have become topic of a rising number of observational, theoretical and laboratory studies. A clear focus has been on amino acids, specifically the simplest amino acid, glycine. Despite theoretical studies and laboratory based work that show that glycine as well as several other amino-acids should form in space (Muñoz Caro et al. 2002, Blagojevic et al. 2003, Congiu et al 2012, Garrod et al. 2013) unambiguous detections are still lacking (Snyder et al. 2005). The search

---

for two other classes of prebiotic compounds – aldoses (polyhydroxy aldehydes) and polyols – has been more successful. Aldoses are compounds with chemical formula $(CH_2O)_n$ containing one aldehyde (-CHO) group. Well-known members of this series are the simple sugars (monosaccharides) glycose, ribose and erythrose. The simplest representative of this class – glycoladehyde ($HC(O)CH_2OH$) – has been successfully detected towards the solar mass protostar IRAS 16293-2422 by ALMA, see Jørgensen et al. (2012) and toward two other objects, the Galactic center source SgrB2(N) (Hollis et al. 2000) and high-mass hot molecular core G31.41+0.31 (Beltran et al. 2009). The best-known representative of the polyols series is glycerine – a basic compound of fats. Glycerine is a triol and has not been detected in space so far, but ethylen glycol ($H_2C(OH)CH_2OH$), a diol, has been observed toward the Galactic center source SgrB2(N) (Hollis et al. 2002) as well as around the low-mass Class 0 protostar NGC 1333-IRAS2A (Maury et al. 2014).

From the chemical structure of both aldoses and polyols it is expected that the key stage in the formation of both molecule classes must be the formation of a chain of carbon-carbon bonds. Furthermore each carbon atom in this chain is attached to an oxygen atom resulting in a -C(O)-C(O)-(C(O))$_n$- backbone. So the crucial stage in the mechanism describing the formation of both $HC(O)CH_2OH$ and $H_2C(OH)CH_2OH$ and other aldoses or polyols not detected in space so far, is the formation of a -C(O)-C(O)- bond. It is here that surface chemistry can play a role. It is generally accepted that carbon-bearing species like $CH_3OH$, $H_2CO$, $CO_2$, and possibly $CH_4$ form on icy dust grains, as these provide surfaces on which gas-phase species accrete, meet, and react. Moreover, these icy grains can absorb excess energy released in a chemical reaction, effectively speeding up processes. Therefore, in dense cold clouds, icy dust grains act both as a molecular reservoir and as a solid-state catalyst. Charnley et al. (2001) suggested that the formation of both glycolaldehyde and ethylen glycol proceeds through the following reaction chain:

$$CO + H \rightarrow HCO, \qquad (1)$$
$$HCO + C \rightarrow HCCO, \qquad (2)$$
$$HCCO + O \rightarrow HC(O)CO, \qquad (3)$$
$$HC(O)CO + H \rightarrow HC(O)CHO, \qquad (4)$$
$$HC(O)CHO + H \rightarrow HC(O)CH_2O / HC(O)CHOH, \qquad (5)$$
$$HC(O)CH_2O / HC(O)CHOH + H \rightarrow HC(O)CH_2OH, \qquad (6)$$
$$HC(O)CH_2OH + H \rightarrow H_2C(O)CH_2OH / HC(OH)CH_2OH, \qquad (7)$$
$$H_2C(O)CH_2OH / HC(OH)CH_2OH + H \rightarrow H_2C(OH)CH_2OH. \qquad (8)$$

Another formation route suggested later by Charnley & Rodgers (2005) is very similar but with altered sequences of H- and O-atom additions. The reduction reactions (1), (5)-(8) (or similar reactions) are reported to proceed under cold molecular cloud conditions (Hiraoka et al 1994, Zhitnikov et al. 2002, Watanabe et al. 2002, Bisschop et al. 2007, Fuchs et al. 2009). The reactions (2) and (3), however, have not been verified experimentally. Furthermore, the formation of a carbon chain consisting of free -C(O)- segments will require the addition of another carbon atom to the final product of reactions (1-4) followed by an O-atom addition or alternatively will require the addition of a C-atom to the product of reaction (2) followed by two place selective oxygen atoms additions. These channels are not considered as effective pathways to form a triple -C(O)- chain.

A considerably more realistic scenario involves reactions between two HCO radicals as produced in the reaction (1) yielding Glyoxal:

$$HCO + HCO \rightarrow HC(O)CHO. \qquad (9)$$

Sequential hydrogenation by 2 or 4 H-atoms turns HC(O)CHO into glycoladehyde and ethylen glycol, respectively, i.e., reactions (2)-(4) in this scheme are replaced by reaction (9), see Fig. **1** for comparison of both schemes. This reaction route may be more relevant than the mechanism suggested by Charnley et al. (2001), as: (i) CO-molecules and H-atoms are among the most abundant species in dense molecular clouds and CO is the second most abundant molecule in interstellar ices; (ii) the formation of $CH_3OH$ proceeds through sequential hydrogenation of CO molecules with HCO as a necessary intermediate which guarantees that this radical is formed on the surface of interstellar grains; (iii) reaction (9) is a radical-radical recombination and expected to be barrierless which is particularly important for the very low temperatures (~10-15 K) of icy grains in space; (iv) the option for further growth of a $(-C(O)-)_n$ backbone through the addition of another HCO radical remains in the case that an H-atom addition in reaction (5) or (7) takes place on the O-atom instead of the C-atom. The dimerisation of HCO, reaction (9) is considered in a recent work by Woods et al. (2013) as one of the formation routes of glycolaldehyde. Their astrophysical model matches the observed estimates in the hot molecular core G31.41+0.31 and low-mass binary protostar IRAS 1629-32422.

In this work, we experimentally investigate surface hydrogenation of CO molecules at dense molecular cloud conditions with the goal to verify the formation of side products of methanol with more than one carbon atom. We will show that for our experimental settings, the formation of glycolaldehyde and ethylen glycol indeed can be realized. This is the first time that a regular hydrogenation scheme of CO is found to result in such complex species. In the discussion section we suggest that the key reaction in the formation of these species involves a HCO + HCO recombination. We also discuss the possible formation of methyl formate ($HC(O)OCH_3$) that is not found in this study but may be formed at other experimental conditions. Subsequently, the experimental results of this work are implemented into a model used by Cuppen et al. (2009) based on the continuous-time, random walk Monte-Carlo method. This model allows for the simulation of microscopic grain-surface chemistry for the long timescales typical in interstellar space, including the layering of ice during the CO freeze out. The choice of microscopic simulations is because HCO radicals have to stay in close vicinity to each other in order for a reaction to occur. Therefore, a method that accounts for the position of the species in the ice lattice is required for the best representation of the system. The astronomical implications are also discussed. First, an overview on the experimental procedure and the performed experiments is given.

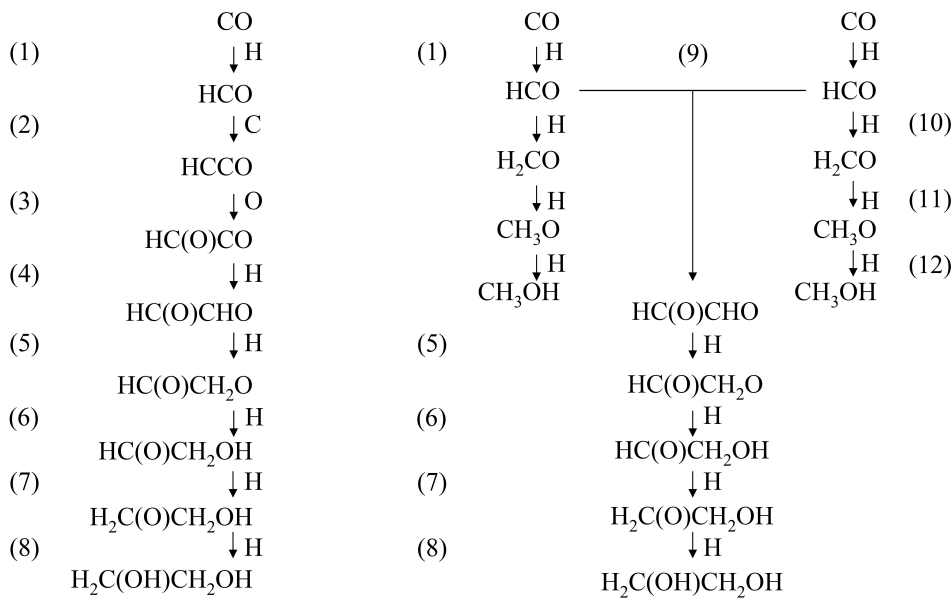

**Figure 1.** A schematic representation of glycolaldehyde and ethylene glycol formation pathways suggested by Charnley et al. (2001) (left scheme) and in this study (right scheme).

**2 Experimental setup**

Experiments are performed using SURFRESIDE$^2$, an ultra-high vacuum (UHV) setup that is described in detail by [Ioppolo et al. (2013)](). This setup comprises three UHV chambers; a main chamber with a base pressure ~ $10^{-10}$ mbar and two chambers housing atom beam lines with base pressures in the range $10^{-10}$-$10^{-9}$ mbar. In the main chamber, a sample holder is mounted on the tip of the cold head of a close-cycle He cryostat. The deposition temperature can be controlled between 13 and 300 K with an absolute precision of <2 K and relative precision of about 0.5 K. Two different atom beam lines are used. A Hydrogen Atom Beam Source (HABS, Dr. Eberl MBE-Komponenten GmbH, see [Tschersich 2000]()) produces atoms by thermal cracking of parent molecules passing through the hot tungsten capillary, while a Microwave Atom Source (MWAS, Oxford Scientific Ltd, see [Anton et al. 2000]()) generates atoms by cracking their parent molecules in a capacity coupled microwave discharge (175 W at 2.45 GHz). In both cases, a nose-shaped quartz pipe is placed along the path of the atom beam to efficiently quench excited electronic or ro-vibrational states of newly formed atoms and non-dissociated molecules through collisions with the walls of the pipe before these reach the ice sample.

The reactivity of CO with H atoms is investigated systematically at 13 K with the goal to verify the formation of C-C bonds. Experiments are performed using co-deposition, i.e., CO-molecules are co-deposited simultaneously with H atoms with pre-defined deposition rates. This allows overcoming the main problem of a sequential deposition technique as used before to study formaldehyde and methanol formation upon CO hydrogenation, i.e., low final yields of the products of CO-hydrogenation due to the limited penetration depth of H-atoms into the pre-deposited CO ice ([Watanabe et al. 2003](), [Fuchs et al. 2009]()). Using a co-deposition technique with an overabundance of H-atoms over CO-molecules makes that virtually all deposited CO molecules are available for hydrogenation reactions and thick (6-30 monolayers) ice of CO-hydrogenation products can be grown. Furthermore, a co-deposition experiment is a more realistic representation of the conditions at which CO is hydrogenated in dense molecular clouds where simultaneous accretion of CO molecules and H-atoms take place but on much longer timescales (see [Cuppen et al. 2009]()).

The newly formed species are monitored *in situ* during co-deposition by means of Fourier Transform Reflection Absorption Infrared Spectroscopy (FT-RAIRS). After finishing the co-deposition, a temperature programmed desorption (TPD) of the ice is routinely performed using quadrupole mass-spectrometry (QMS) to monitor in the gas-phase thermally desorbing species. The use of TPD as main analytical tool is necessary since the strong IR absorption features of glycoladehyde (HC(O)CH$_2$OH), ethylen glycol (H$_2$C(OH)CH$_2$OH) and also glyoxal (HC(O)CHO) overlap with the absorption bands of H$_2$CO and CH$_3$OH, making unambiguous spectroscopic assignments challenging (see also [Öberg et al. 2009]()).

Several control experiments are performed to confirm that the formation of glycoladehyde and ethylene glycol takes place indeed at 13 K and that this is not the result of thermal processing, recombination of trapped radicals during the TPD, gas-phase reactions or contaminations in the atom beam lines. Isotope shift experiments using $^{13}$CO instead of $^{12}$CO or D instead of H are performed to further constrain the results. All experiments are summarized in Table 1. The numbering in the first column is used for cross-referencing.

Table 1. Overview of performed experiments.

| | Experiment | $T_{sample}$, K | Ratio | $CO_{flux}$, cm$^{-2}$s$^{-1}$ | $H_{flux}$, cm$^{-2}$s$^{-1}$ | t, min | TPD | Detection[c] |
|---|---|---|---|---|---|---|---|---|
| 1 | CO + H | 13 | 1:5 | 1.5E12 | 8E12 | 360 | QMS$^{2K/10K}$ | Y |
| 2 | CO + H | 13 | 1:5 | 1.5E12 | 8E12 | 360 | [b]RAIRS$^{2K}$ | Y |
| 3 | CO + H | 25 | 1:5 | 1.5E12 | 8E12 | 360 | QMS$^{2K/10K}$ | N |
| 4 | CO + H | 13 | 1:25 | 3E11 | 8E12 | 360 | QMS$^{2K/10K}$ | Y |
| 5 | CO + H[a] | 13 | 1:25[a] | 3E11 | 8E12[a] | 360 | QMS$^{2K/10K}$ | Y |
| 6 | CO + D | 13 | 1:25 | 3E11 | 8E12 | 360 | QMS$^{2K/10K}$ | N |
| 7 | $^{13}$CO + H | 13 | 1:25 | 3E11 | 8E12 | 360 | QMS$^{2K/10K}$ | Y |
| 8 | CO + H | 13 | 1:25 | 3E11 | 8E12 | 72 | QMS$^{2K/10K}$ | Y |
| 9 | CO + H | 13 | 1:25 | 3E11 | 8E12 | 360 | QMS$^{5K}$ | Y |
| 10 | CO + H | 13 | 1:25 | 3E11 | 8E12 | 600 | RAIRS$^{5K}$ | Y |
| | | | | $CH_3OH_{flux}$, cm$^{-2}$s$^{-1}$ | $H/H_2$ $_{flux}$, cm$^{-2}$s$^{-1}$ | | | |
| 11 | CH$_3$OH+H | 13 | 1:25 | 3E11 | 8E12 | 360 | QMS$^{5K}$ | N |
| 12 | CH$_3$OH+H$_2$ | 13 | 1:25 | 3E11 | - | 360 | QMS$^{5K}$ | N |

Experiments are performed using co-deposition technique; $X_{flux}$ is the deposition rate of a selected species expressed in particles per cm$^2$ per s, $T_{sample}$ is the substrate temperature during co-deposition; *t* is the time of co-deposition; *TPD* is the temperature programmed desorption experiment performed afterward with the TPD rate indicated, normally the ice is gently warmed up to remove the bulk of CO, then high TPD rate is used to increase the sensitivity of a technique, *Detection* indicates whether glycoladehyde and ethylene glycol are identified. [a]The microwave discharge is used to generate the H-atom beam instead of a thermal cracking source. [b]Instead of a TPD with a constant rate, annealing at a number of chosen temperatures is made with a simultaneous RAIR spectra recording. [c]detection of both glycoladehyde and ethylene glycol in the experiment

## 3 Results

In Fig. **2**, a typical example of a QMS TPD spectrum is presented for a CO + H experiment at 13 K (exp. 4 in Table 1). One can see that in the temperature range from 50 to 250 K, four CO hydrogenation products show up. In addition to peaks originating from the previously detected H$_2$CO (~ 100 K) and CH$_3$OH (~ 140 K) (Hiraoka et al 1994, Zhitnikov et al. 2002, Watanabe et al. 2002, Fuchs et al. 2009), there are two more desorption peaks, one centered at 160 K and one at 200 K. The higher desorption temperatures w.r.t. the values for H$_2$CO (100 K) and CH$_3$OH (140 K) are consistent with less volatile and heavier carriers. The TPD QMS spectra provide information to identify the origin of these carriers; molecules desorb at specific temperatures and fragmentation patterns upon electron impact induced dissociative ionization are available for different electron energies for many different species. The potential of this method – linking two different physical properties - is illustrated for the H$_2$CO and CH$_3$OH bands. The desorption bands are linked to mass spectra that can be compared to literature values upon 60 eV electron impact ionization for formaldehyde and methanol, as shown in the two top-left insets in Fig. **2**. The experimental and database fragmentation patterns are very similar. Small inconsistencies are likely due to partial co-desorption of CH$_3$OH with H$_2$CO at 100 K and *vice versa* co-desorption of H$_2$CO trapped in the bulk of CH$_3$OH at 140 K.

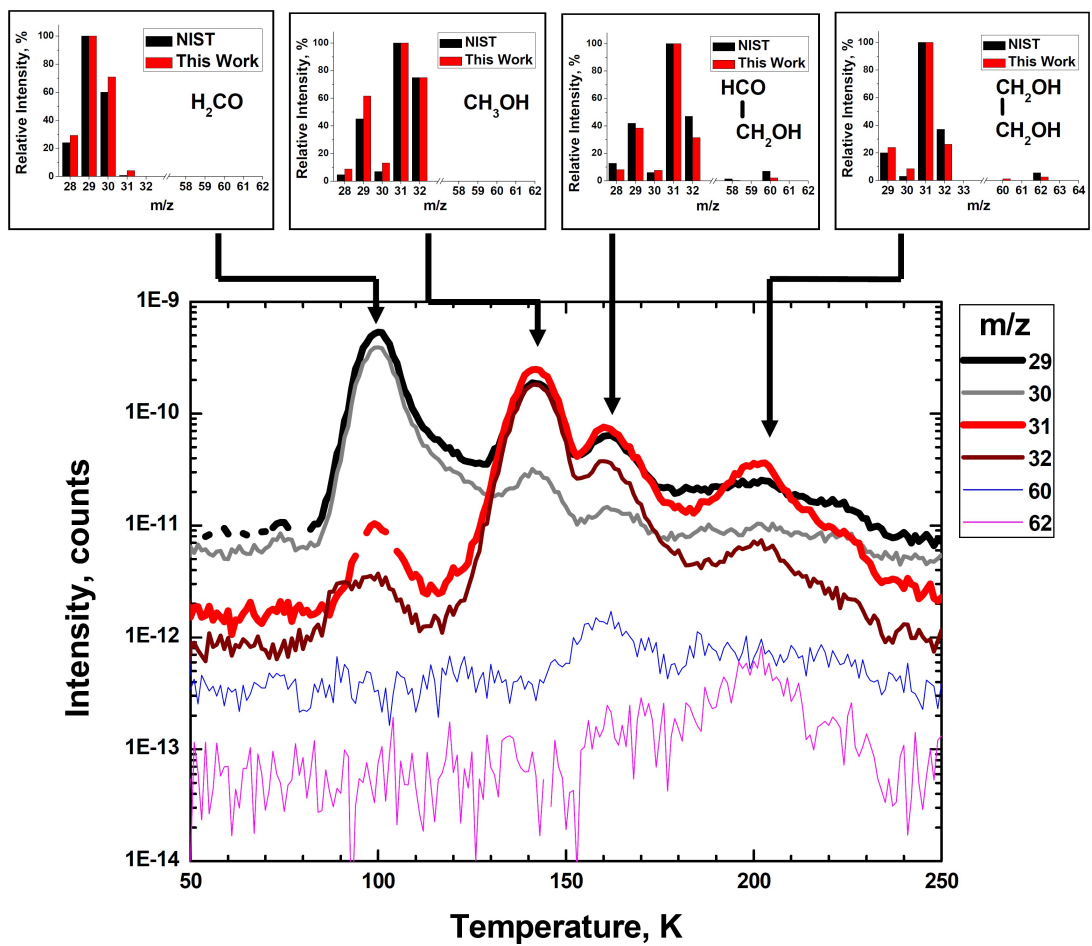

**Figure 2.** TPD spectra obtained after experiment 4 (Table 1) for the indicated m/z numbers. Insets on top of the figure compare the fragmentation patterns of desorbing species detected in this experiment upon 60 eV electron impact with those available from literature[1].

In a similar way, the two additional desorption peaks at 160 K and 200 K can be assigned to glycoladehyde and ethylene glycol, respectively. The desorption temperatures of these species are consistent with the values available from Öberg et al. (2009) (see their Fig. **12**, where $CH_3OH$, $HC(O)CH_2OH$, and $H_2C(OH)CH_2OH$ desorption peaks are reported to be at 130, 145, and 185 K, respectively). Moreover, the observed fragmentation pattern upon 60 eV electron-impact ionization is very similar to the literature values, as illustrated in the two top-right insets of Fig. **2**.

---

[1] NIST Mass Spec Data Center, S. E. Stein, director, "Mass Spectra" in NIST Chemistry WebBook, NIST Standard Reference Database Number 69, Eds. P.J. Linstrom and W.G. Mallard, National Institute of Standards and Technology, Gaithersburg MD, 20899

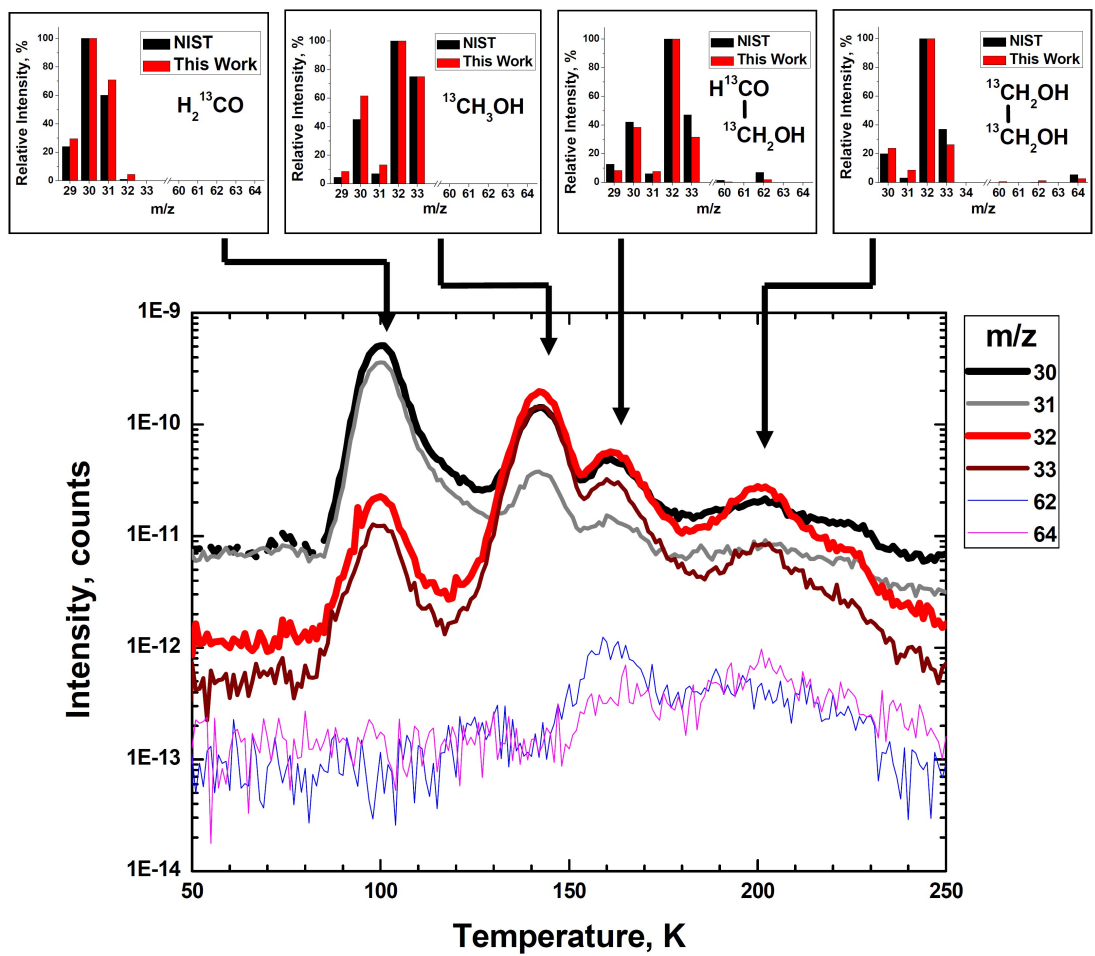

**Figure 3.** TPD spectra obtained after experiment 7 (Table 1) for the indicated m/z numbers. Insets on top of the figure show the comparison between fragmentation patterns of the desorbing species detected in this experiment upon 60 eV electron impact with those available from literature.

This assignment can be further constrained by performing the same experiments using isotopes. Deuteration experiments are not ideal for this. Hidaka et al. (2007) found that the CO deuteration rate at 15 K is 12.5 times lower than the corresponding value for hydrogenation. Our experimental results confirm this finding and show a substantial decrease for D-containing products. The $D_2CO$ and $CD_3OD$ formation yields are roughly 10 times lower and the $DC(O)CD_2OD$ and $D_2C(OD)CD_2OD$ signals are below the detection limits. Therefore, $^{13}CO$ experiments are used to provide additional proof for glycolaldehyde and ethylene glycol formation at our experimental settings. Moreover, with $^{13}CO$ as precursor, dissociative ionization products containing only one carbon atom will be shifted by one m/z number, while fragments containing two carbon atoms will shift by two. This provides a tool to identify unambiguously C-C bond formations. In Fig. **3**, the result of co-deposition of H atoms and $^{13}CO$ molecules is shown, for conditions that are nearly identical to those applied in Fig. **2**. Again, four peaks are found in the TPD QMS spectra. The insets in the top-side of Fig. **3** compare the fragmentation patterns of desorbing species recorded in our experiment with values obtained by extrapolating literature results. This extrapolation is realized by adding m/z=1 to the masses from 28 to 33 and m/z=2 to the masses from 58 to 64 for the fragmentation pattern of regular $^{12}C$ glycoladehyde and ethylene glycol, i.e., according to the number of carbon atoms in the corresponding ions. Again the experimental and literature values are very close and fully consistent with the previous finding that glycolaldehyde ($H^{13}C(O)^{13}CH_2OH$) and ethylen glycol ($H_2^{13}C(OH)^{13}CH_2OH$) form along with the formation of $H_2^{13}CO$ and $^{13}CH_3OH$ upon surface hydrogenation of CO molecules.

Repeating experiment 4 depicted in Fig. **2** for a co-deposition time about twice as long, using RAIRS instead of the QMS allows us to obtain RAIR difference spectra of the two ice constituents desorbing at about 160 and 200 K. The differences are determined by subtracting spectra at 152 and 168 K (Fig. **4**a), and 183 and 210 K (Fig. **4**b), i.e., before and after desorption of the two individual TPD features. Low peak-to-noise ratios complicate the identification of absorption features, but tentative assignments can be made. The three strongest absorption features of $HC(O)CH_2OH$ are visible in Fig. **4**a, i.e., the OH-stretch mode in the range from 3600 to 3000 $cm^{-1}$, coinciding CH and $CH_2$ stretching modes in the range from 3000 to 2800 $cm^{-1}$ and the sharp CO stretching mode at 1750 $cm^{-1}$. Similarly, both OH stretching and $CH_2$ stretching modes of $H_2C(OH)CH_2OH$ (its strongest absorption features) are visible in Fig. **4**b, while the absorption feature in the range 1400-1500 $cm^{-1}$ is likely due to $CH_2$ scissor and OH bending modes. Unfortunately, C-C stretching mode of $H_2C(OH)CH_2OH$ lies within the range 1000-1100 $cm^{-1}$ that can not be observed due to the artifact caused by difference in the shape of background spectra depending on the temperature of the sample. Furthermore, it overlaps with one of the strongest absorption features of $CH_3OH$. The C-C stretching mode of $HC(O)CH_2OH$ is expected to be found around 870 $cm^{-1}$ and cannot be assigned due to the low peak-to-noise ratio in this region of the spectra (Buckley & Giguère 1966, Kobayashi et al. 1976, Hudson et al. 2005, Ceponkus et al. 2010).

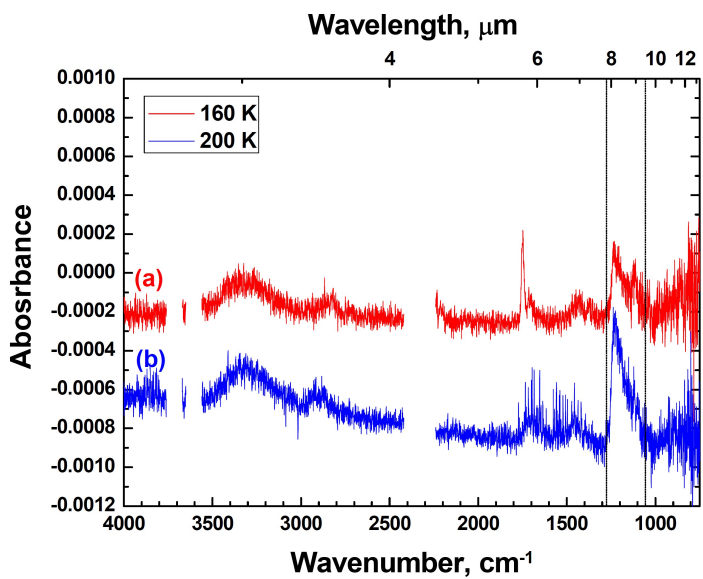

**Figure 4.** RAIR difference spectra obtained between a) 153 and 168 K, and b) 183 and 210 K in experiment 10. The sharp absorption feature in the range between 1270 and 1050 $cm^{-1}$ is an artifact caused by the difference in the shape of background spectra depending on the temperature of the sample. Cuts between 2250-2400 and 3560-3760 $cm^{-1}$ are because of atmospheric $CO_2$ absorbance along the path of the IR beam outside of the main chamber.

In addition, a number of complementary/control experiments has been performed. An increase of the CO-deposition rate by a factor five (experiment 1), a change of TPD rate or total co-deposition time (experiments 8 and 9), and the use of a microwave discharge source instead of the thermal cracking source to generate the H atoms (experiment 5) all do not qualitatively change the results depicted in Figs. **2** and **3**. No $HC(O)CH_2OH$ and $H_2C(OH)CH_2OH$ formation is found for a co-deposition experiment of H atoms with CO molecules at 25 K (instead of 13 K) while only traces of $H_2CO$ and $CH_3OH$ can be detected (experiment 3). Signatures of glycolaldehyde and ethylene glycol also are not found in $CH_3OH+H$ and $CH_3OH+H_2$ co-deposition experiments performed under similar conditions and applying similar co-deposition rates (experiments 11 and 12).

**4. Discussion**

The experimental results presented in the previous section show that glycol aldehyde and ethylene glycol could be formed in co-deposition experiments of CO molecules and H atoms. This is an important experimental finding, as so far hydrogenation reactions were mainly shown to be effective in the formation of smaller species (e.g., ammonia from N + H) with $CH_3OH$ (CO + H) as the largest species systematically studied so far by more than one independent group. Instead, experimental studies showed that solid-state reactions induced by vacuum UV irradiation, cosmic ray or electron bombardment offer pathways to form molecules with up to 10-12 atoms. In Öberg et al. (2009), for example, VUV irradiation of a pure methanol ice was shown to result in the formation of both glycolaldehyde and ethylene glycol. The formation scheme presented here does not require energetic processing and can proceed at 13 K. The work discussed here is an extension of the well-studied formaldehyde and methanol (CO+H) formation scheme that is generally accepted as the dominant pathway explaining the observed large $CH_3OH$ abundances in space. Therefore, this process should also be efficient at cold dense clouds conditions, particularly during the CO freeze-out stage that takes place well before radiation from a newly formed protostar becomes important. To which extent, however, the different phases (read processes) in star formation determine glycolaldehyde and ethylene glycol abundances is hard to estimate. Detection of both $HC(O)CH_2OH$ and $H_2C(OH)CH_2OH$ in our experiments as well as a gradual decrease in abundances along the order $H_2CO>CH_3OH>HC(O)CH_2OH>H_2C(OH)CH_2OH$ is consistent with the mechanism proposed in the present work, i.e., a sequence of surface reactions (1), (9), and (5)-(8).

The only experimental observation that seems to be contradicting our interpretation is the experimental non-detection of glyoxal (HC(O)CHO). A possible explanation is that reaction (5) has no or a very low activation barrier, compared to that of reactions (1), (7) and (11):

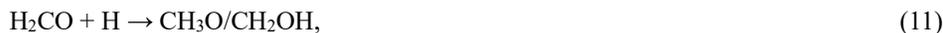

$$H_2CO + H \rightarrow CH_3O/CH_2OH, \hspace{4cm} (11)$$

that are the limiting steps in the formation of $H_2CO$, $H_2C(OH)CH_2OH$, and $CH_3OH$, respectively. Unfortunately, there exists no experimental data for the activation barrier of reaction (5). Galano et al. (2004) performed quantum chemical calculations of the interaction of glyoxal with OH radicals and concluded that the OH addition and formation of an intermediate complex followed by H-atom abstraction proceeds barrierless. This is consistent with experimental gas-phase results of Feierabend et al. (2008), who found that H-atom abstraction from glyoxal by OH radicals has a negative temperature dependence with a slight deviation from Arrhenius behavior that is reproduced over the temperature range 210 – 396 K. However, it contradicts the results by Woods et al. (2013) who find a barrier of 1108 K in their calculations. Although this activation barrier appears rather high, it can easily be overcome if tunneling is involved, e.g., the CO + H barrier is roughly three times higher. In this case we would not necessarily expect to detect glyoxal along with $HC(O)CH_2OH$ and $H_2C(OH)CH_2OH$ for our settings. Therefore, it seems correct to conclude that the interaction of an H atom with glyoxal proceeds very efficiently, barrierless or through tunneling, followed by an H-atom addition instead of abstraction, consistent with the non-detection of glyoxal in our experiments.

Other possible scenarios leading to the formation of C-C bonds that are mentioned in literature are:

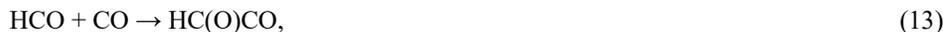

$$HCO + CO \rightarrow HC(O)CO, \hspace{4cm} (13)$$

followed by reaction (4) to yield HC(O)CHO, and

$$HCO + H_2CO \rightarrow HC(O)CH_2O. \tag{14}$$

Reaction (13) should possess an activation barrier and leads again to the glyoxal formation, i.e., this means that the assumption of a barrierless hydrogenation of glyoxal, as made above, still should hold to explain the observed results. Reaction (14) indeed helps to bypass the glyoxal formation as an intermediate and leads directly to the formation of the observed glycolaldehyde through reaction (6). However, this reaction also expected to possess an activation barrier and is not reported in literature. Reaction (9), therefore, is proposed as the key step responsible for the formation of C-C bonds and we assume that the hydrogenation of glyoxal proceeds barrierless.

We would like to stress that both glycolaldehyde and ethylene glycol are formed *in situ* at 13 K and are not the result of the recombination of the formed HCO and $CH_3O/CH_2OH$ radicals during the TPD. This is because: (i) in the case that trapped radicals do recombine, glyoxal (as the product of HCO dimerisation) should be visible in our experiments (as discussed before, this is not the case); (ii) these radicals are not observed by RAIRS; and iii) a lack of a qualitative difference between experiment 4 and experiments 8 and 9 (where different TPD rates and exposure times are used) is consistent with the conclusion that both glycolaldehyde and ethylene glycol are not the result of thermally induced chemistry.

Moreover, up to our knowledge, there exist no systematic studies for $CO:H_2CO:CH_3OH$ mixtures, while thermally induced chemistry of pure $CH_3OH$ (experiment 12) and $H_2CO$ (Schutte et al. 1993, Noble et al. 2012) do not result in the formation of $HC(O)CH_2OH$ or substantial amounts of $H_2C(OH)CH_2OH$. Furthermore, in experiments where pure $H_2CO$ and $CH_3OH$ are first irradiated by UV-photons and then analyzed by means of TPD the formation of glycolaldehyde or ethylene glycol are not reported either, while the methyl formate ($HC(O)OCH_3$) is clearly observed (Gerakines et al. 1996). The non-detection of abundant $HCOOCH_3$ in our experiments confirms that our results are due to non-energetic processing.

Another point that has to be noted is that under our experimental conditions, the $H_2CO$ yields dominate over the $CH_3OH$ yields, and this then applies to intermediate products of $H_2CO$ hydrogenation through reaction (11) that are expected to be less abundant than HCO – a product of CO hydrogenation. Therefore, we expect that $CH_3O$ and $CH_2OH$ are not abundant in our experiments and will not contribute significantly to the reactions that lead to the formation of complex species.

However, subsequent interactions of $CH_3O$ or $CH_2OH$ with HCO radicals may lead to the formation of methyl formate ($HC(O)OCH_3$) and again glycolaldehyde:

$$HCO + CH_3O \rightarrow HC(O)OCH_3, \tag{15}$$
$$HCO + CH_2OH \rightarrow HC(O)CH_2OH. \tag{16}$$

Only traces of the methyl formate are found in experiment 1 (Table 1), and these are actually within the experimental uncertainty. In the future, another set of experiments with a significantly higher hydrogenation degree of CO may verify the possibility of reactions (15) and (16) to occur, but for the moment we consider this outside the scope of the present work.

## 5. Astrochemical implications

The astrochemical importance of the experimental findings discussed here is that the formation of complex molecules can occur already in the dark ages of star formation, i.e., in a period when energetic processing is expected not to be very relevant. The focus here has been on two important species: glycolaldehyde is often considered to be the simplest monosaccharide, and ethylene glycol is the first representative of the polyol family, of which the triol glycerine is the most known one.

The solid state reaction schemes at play have to be known to understand where molecules of prebiotic interest are likely to be formed. To investigate the possible astrochemical relevance of the laboratory findings presented here we have implemented the new suggested reaction route (1), (9), (5)-(8) into a model previously used to simulate the formation of $H_2CO$ and $CH_3OH$ on interstellar ice surfaces. This model utilizes the continuous-time random-walk Monte Carlo method, which simulates microscopic grain-surface chemistry for timescales as typical for the interstellar medium. This model has been described in detail in Fuchs et al. (2009) and Cuppen et al. (2009), and for additional information the reader is referred to these papers. The model simulates a sequence of processes that can occur on a grain surface. This grain surface is modeled as a lattice with the number of sites determined by the size of the grain and the site density for the adsorbing CO. The order of this sequence is determined by means of a random number generator in combination with the rates for the different processes. These processes include deposition onto the surface, hopping from one lattice site to a nearest neighbor, desorption of the surface species, and reactions between two species. Each of these processes is characterized by a specific rate through an activation energy barrier. Here we use reaction activation barriers and energy parameters determining hopping and desorption activation barriers as derived in Fuchs et al. (2009). In order to incorporate formation routes to the newly observed species, the model is extended and five new reactions are incorporated in the chemical network. Reactions (9), (6), and (8) are set to be barrierless as radical-radical recombination reactions, reaction (5) is also set to be barrierless (see aforementioned discussion), while reaction (7) is set to have an activation energy barrier of the same value as the comparable reaction (11) obtained by Fuchs et al. (2009). The reaction rate coefficients used in the simulations are summarized in Table 2. Consequently, five new species, including glycolaldehyde and ethylene glycol, need to be introduced in the simulations and these are $HC(O)CHO$, $HC(O)CH_2O/HC(O)CHOH$, $HC(O)CH_2OH$, $H_2C(O)CH_2OH/HC(OH)CH_2OH$ and $H_2C(OH)CH_2OH$. Since all of them are heavy, low-volatile species, the energy parameter $E$ responsible for the hopping and desorption of species is set to the same value as for $CH_3OH$, effectively immobilizing these species.

Table 2. A list with the reaction rate coefficients, $R_{react}$, for the key reactions used in the simulations at 12 and 16.5 K grain temperatures (see Fuchs et al. 2009 for more details).

| N | Reaction | $R_{react}$ (s$^{-1}$) for 12 K | $R_{react}$ (s$^{-1}$) for 16.5 K | N | Reaction | $R_{react}$ (s$^{-1}$) for 12 K | $R_{react}$ (s$^{-1}$) for 16.5 K |
|---|---|---|---|---|---|---|---|
| (1) | $CO + H$ | $2 \cdot 10^{-3}$ | $4 \cdot 10^{-3}$ | (9) | $HCO + HCO$ | $2 \cdot 10^{11}$ | $2 \cdot 10^{11}$ |
| (10) | $HCO + H$ | $2 \cdot 10^{11}$ | $2 \cdot 10^{11}$ | (5) | $HC(O)CHO + H$ | $2 \cdot 10^{11}$ | $2 \cdot 10^{11}$ |
| (11) | $H_2CO + H$ | $2 \cdot 10^{-4}$ | $2 \cdot 10^{-2}$ | (6) | $HC(O)CH_2O/ HC(O)CHOH + H$ | $2 \cdot 10^{11}$ | $2 \cdot 10^{11}$ |
| (12) | $CH_3O/CH_2OH + H$ | $2 \cdot 10^{11}$ | $2 \cdot 10^{11}$ | | | | |
| | | | | (7) | $HC(O)CH_2OH + H$ | $2 \cdot 10^{-4}$ | $2 \cdot 10^{-2}$ |
| | | | | (8) | $H_2C(O)CH_2OH/ HC(OH)CH_2OH + H$ | $2 \cdot 10^{11}$ | $2 \cdot 10^{11}$ |

In Fig. **5**, the outcome of four different test simulations is presented. The lower panels show simulations using identical input parameters as for the simulations presented in the lower panels of Fig. **3** in Cuppen et al. (2009), but with the four new reaction routes added to the code. These parameters are $n_H = 1 \cdot 10^5$ cm$^{-3}$, $n_{grain} = 1 \cdot 10^{-12} n_H$ and the gas-phase CO initial abundance equals to $n(CO)_{initial} = 1 \cdot 10^{-4} n_H$, i.e. 10 cm$^{-3}$. In these two runs, a high value for the hydrogen atom density $n(H) = 10$ cm$^{-3}$ (Goldsmith & Li 2005) is used as described in Cuppen et al. (2009). The corresponding grain temperatures are 12 and 16.5 K for panels c and d, respectively. In the upper panel the results of the same simulations are presented, but here the H-atom density is set 10 times less than in the simulations presented in the lower panel of Fig. **5** and equals $n(H) = 1$ cm$^{-3}$ (Duley & Williams 1984). This represents the low H-atom density case. All results are converted to grains with a standard size of 0.1 μm. The choice of $n(H)$ to be 10 cm$^{-3}$ and 1 cm$^{-3}$ for high- and low-density cases, respectively, reflects two extreme conditions; i) when most of the CO will be hydrogenated to the final product, i.e., $CH_3OH$, and ii) most of the CO will be locked in a non-hydrogenated state. A more realistic $n(H)$ value is likely closer to 2-3 cm$^{-3}$, and is covered within the two boundary conditions discussed here.

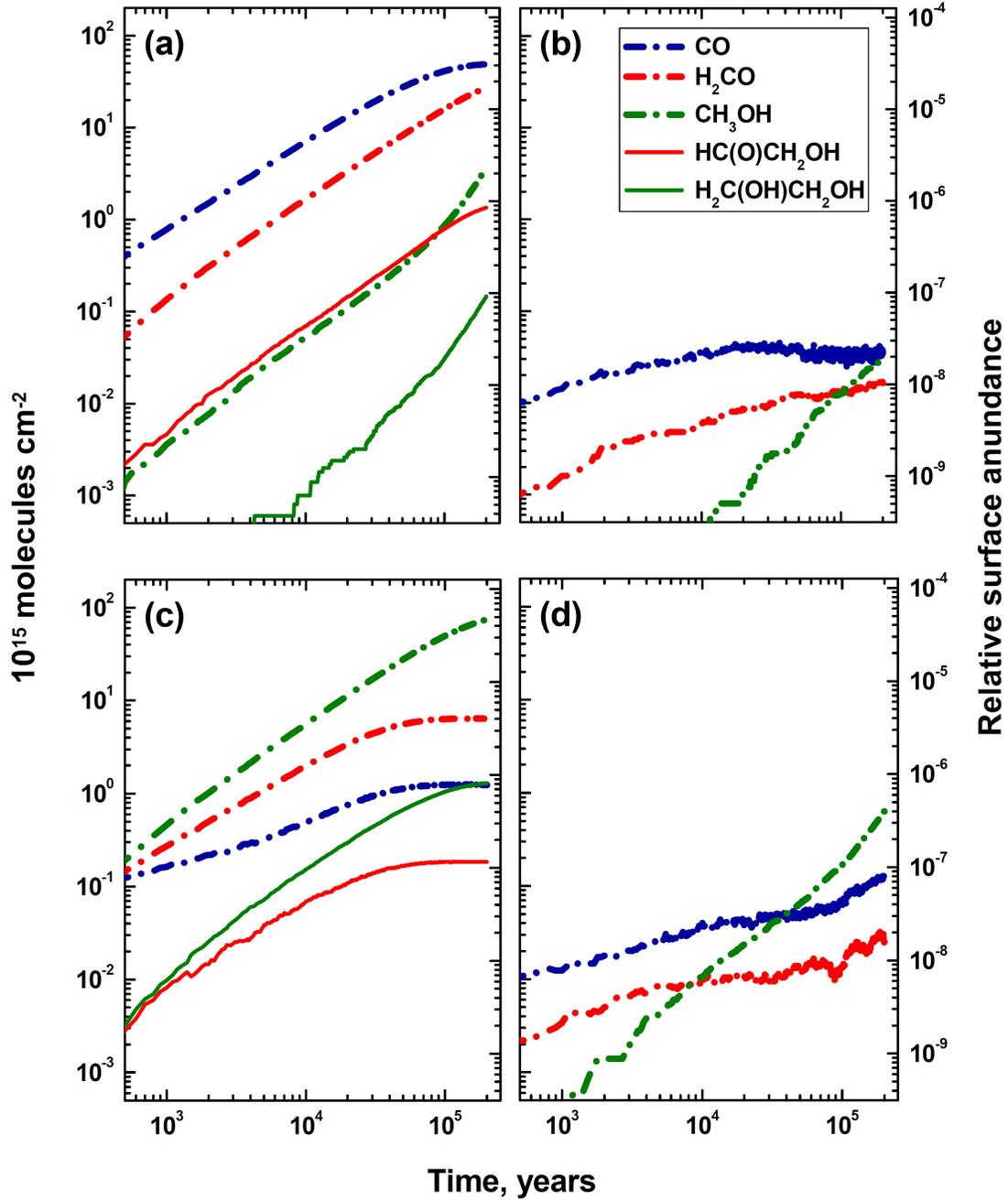

**Figure 5.** CO, H$_2$CO, CH$_3$OH, HC(O)CH$_2$OH and H$_2$C(OH)CH$_2$OH build up as a function of time for $n_H = 1 \cdot 10^5$ cm$^{-3}$, $n_{grain} = 1 \cdot 10^{-12} n_H$ and $n(CO)_{initial} = 1 \cdot 10^{-4} n_H$. (a) T = 12 K and $n(H) = 1$ cm$^{-3}$ (b) T = 16.5 K and $n(H) = 1$ cm$^{-3}$ (c) T = 12 K and $n(H) = 10$ cm$^{-3}$ (d) T = 16.5 K and $n(H) = 10$ cm$^{-3}$. The relative surface abundance is given with respect to $n_H$.

The CO, H$_2$CO and CH$_3$OH abundances shown in the two lower panels of Fig. 5 follow the same trends and similar abundances (for the same input parameters) as in Cuppen et al (2009) (see lower panel of their Fig. 3). This is also expected. The upper panels show the H atom low-density case that was not discussed in Cuppen et al (2009). A further extension is realized by inclusion of the five new reaction routes that are incorporated in our code and that aim at introducing glycolaldehyde and ethylene glycol formation in interstellar reaction schemes, following the experimental confirmation discussed in the previous section of this paper. Therefore, two additional curves are presented in panels a and c that show the evolution of the HC(O)CH$_2$OH and H$_2$C(OH)CH$_2$OH abundances, respectively.

Both abundances experience a clear growth, and the glycolaldehyde abundance correlates with the formaldehyde abundance, while the abundance of ethylene glycol correlates with the abundance of methanol. This is not surprising since both CH$_3$OH and H$_2$C(OH)CH$_2$OH are hydrogen saturated species while both H$_2$CO and HC(O)CH$_2$OH are not. The HC(O)CH$_2$OH/H$_2$CO ratio is kept within a 3-5 % range, while the H$_2$C(OH)CH$_2$OH/CH$_3$OH ratio is about 2-4 %. These relatively high values for HC(O)CH$_2$OH and H$_2$C(OH)CH$_2$OH formation are in fact of the same order as derived from the experiments shown in Figs. **2** and **3**. An exact comparison of the amounts of formed species requires a precise knowledge of ionization cross-sections for all desorbing species as well as their pumping efficiencies, and this information is not available. A rough estimation, however, can be given just by integrating the corresponding areas in the TPD curves. By comparing the QMS TPD areas for all four species, HC(O)CH$_2$OH/H$_2$CO and H$_2$C(OH)CH$_2$OH/CH$_3$OH ratios are found in the range of 1-9% (depending on the m/z number used for comparison), covering the 3-5 and 2-4 % values mentioned above.

It should be noted that the new reaction channels introduced here, leading to the formation of HC(O)CH$_2$OH and H$_2$C(OH)CH$_2$OH do not affect the CO, H$_2$CO, and CH$_3$OH abundances and their dependencies, and, consequently, do not change previous conclusions from Cuppen et al (2009). This is mainly due to the low final amount of glycolaldehyde and ethylene glycol formed in the simulations that lock only a few % of the originally available CO.

In Fig. **6**, the cross sections of the grown ice mantles are shown, similar to Cuppen et al (2009) (see their Figs. **5** and **6**). Correlations between the abundances of glycolaldehyde and formaldehyde and between the abundances of ethylene glycol and methanol find a further confirmation in these plots. Moreover, one can see that the distribution of HC(O)CH$_2$OH (light green) among the cross-section of the ice mantle matches the H$_2$CO (orange) distribution, while H$_2$C(OH)CH$_2$OH (dark green) correlates with the CH$_3$OH (red), further suggesting that these species are chemically linked.

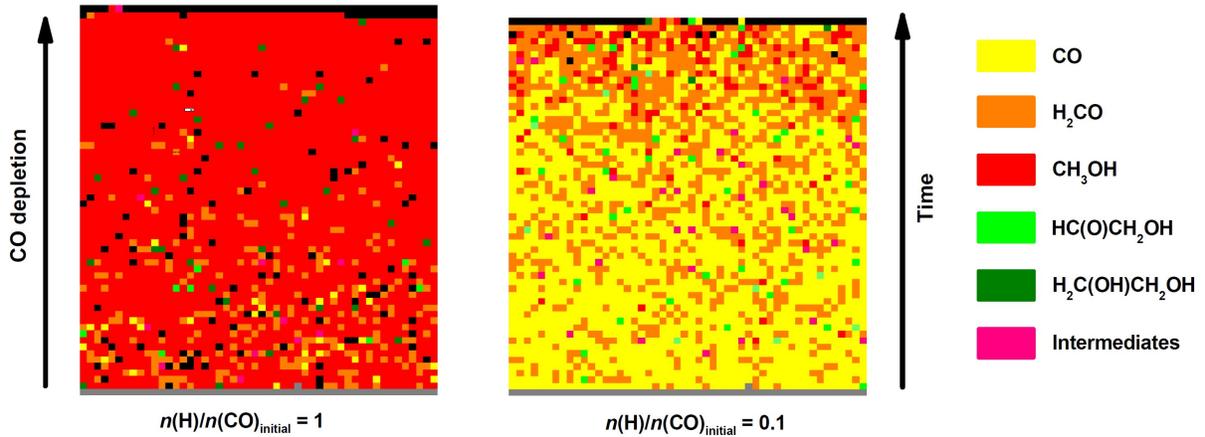

**Figure 6.** Schematic picture of the growth of the ice mantle after $2 \cdot 10^5$ years at $n_H = 1 \cdot 10^5$ cm$^{-3}$, $n_{grain} = 1 \cdot 10^{-12}$ $n_H$ and $n(CO)_{initial} = 1 \cdot 10^{-4}$ $n_H$ for 12 K grain temperature. Left panel $n(H) = 10$ cm$^{-3}$. Right panel $n(H) = 1$ cm$^{-3}$. Grain surface is indicated by brown colour, unoccupied cites by black, CO is yellow, H$_2$CO is orange, CH$_3$OH is red, HC(O)CH$_2$OH is light green, and H$_2$C(OH)CH$_2$OH is dark green. Magenta corresponds to all kinds of intermediate radicals.

As stated before, this combined experimental and theoretical study shows that formation of both glycolaldehyde and ethylene glycol may take place already in the prestellar stage well before energetic processing of the ice by the newly formed protostar will take place. The correlation between the abundances of HC(O)CH$_2$OH and $H_2C(OH)CH_2O$ with those of formaldehyde and methanol, typically in ratios of the order of a few percent, is as expected. $CH_3OH$ is a common component of interstellar ices, and its formation mainly proceeds through sequential surface hydrogenation of CO molecules during the CO-freeze-out stage. We therefore expect that the formation of both $HC(O)CH_2OH$ and $H_2C(OH)CH_2O$ proceeds during the same stage of the molecular cloud evolution as the formation of $CH_3OH$ ice, but with substantially lower abundances.

It is hard to say more on how this relates to complex molecule formation during a later stage, upon energetic processing. UV photo-processing of interstellar ice analogues has been proposed as the way to form prebiotic species in space. Early experiments, pioneered by Hagen, Allamandola, and Greenberg (1979), revealed that such molecules form upon long term vacuum UV exposure of low temperature ices comprising a mixture of astronomically relevant constituents. This discovery has motivated many of the ice chemistry studies that were performed in the following decades, with two key papers published by Bernstein et al. (2002) and Munoz-Caro et al. (2002). In Öberg et al. (2009) the formation of larger complex species, including glycolaldehyde and ethylene glycol have been extensively described upon Lyman-α irradiation of pure methanol ice. In Table 6 of Öberg et al. (2009) a comparison is made for the abundance ratios of complex molecules in the gas phase detected towards a variety of astrophysical environments and the ratios found upon vacuum UV processing of a pure methanol ice.

It is not our intention to discriminate between H-atom addition and UV irradiation processes, or others, like electron bombarded ice as described by Arumainayagam et al. (2010), but to show that solid state reactions as these and studied in the laboratory offer a pathway for complex molecule formation in space. In this paper, this is discussed for the first time for sequential atom addition reactions. Glycolaldehyde has been successfully detected toward the low-mass protostar IRAS 16293-2422 (Jørgensen et al. 2012) with abundances of $6 \cdot 10^{-9}$ relative to $H_2$. This reported gas-phase abundance is well below the abundance of $HC(O)CH_2OH$ obtained in our simulations (see Fig. 5) and therefore consistent with a scenario where it has been sublimated or non-thermally desorbed from the solid state. It is also worth comparing abundance ratios, specifically for species that are considered to be chemically linked (i.e., $HC(O)CH_2OH/H_2CO$ and $H_2C(OH)CH_2OH/CH_3OH$). In a rough model one may assume that solid-state and gas-phase ratios should be comparable, although we stress that different desorption mechanisms – (non) thermal desorption, chemisorption, or even grain collisions – may have different efficiencies for different molecules, specifically over large time scales covering different evolutionary stages. However, as stated above, the goal here is to show that the numbers have the right order of magnitude. Observations toward the same source performed by Schöier et al. (2002) reported a gas-phase $H_2CO$ abundance of $6 \cdot 10^{-8}$ and a solid-state abundance of $(1-4) \cdot 10^{-6}$ relative to $H_2$, yielding a gas-phase $HC(O)CH_2OH/H_2CO$ ratio of 10 %, comparable within the uncertainties to the 3-5 % value found in the simulations and (1-9)% concluded from the experiments. In a similar way, correlations between ethylene glycol and methanol abundances yield $H_2C(OH)CH_2OH/CH_3OH$ ratios of 2-4 % found in the simulations and (1-9)% in the experiments. Jørgensen et al. (2012) give a tentative assignment of $H_2C(OH)CH_2OH$ toward the low-mass protostar IRAS 16293-2422 with a relative abundance of ethylene glycol of 0.3-0.5 with respect to glycolaldehyde, i.e., $(2-3) \cdot 10^{-9}$, while Schöier et al. (2002) for the same source reported $CH_3OH$ abundance of $3 \cdot 10^{-7}$. This results in a $H_2C(OH)CH_2OH/CH_3OH$ ratio of about 1%.

More observational data, specifically in the prestellar phase are needed to link the present hydrogenation laboratory data to solid state efficiencies in space. The important conclusion that stands is that atom addition reactions – in the past specifically proposed as important for the formation of smaller molecules, including $H_2O$ and $NH_3$, also can contribute to molecular complexity in space, i.e., far beyond methanol formation.

## ACKNOWLEDGMENTS

The SLA group has received funding from the European Community's Seventh Framework Programme (FP7/2007-2013) under grant agreement no. 238258, the Netherlands Research School for Astronomy (NOVA) and from the Netherlands Organization for Scientific Research (NWO) through a VICI grant. H.M.C. is grateful for support from the VIDI research program 700.10.427, which is financed by The Netherlands Organisation for Scientific Research (NWO) and from the European Research Council (ERC-2010-StG, Grant Agreement no. 259510-KISMOL). S.I. is supported by the Marie Curie Fellowship (FP7-PEOPLE-2011-IOF-300957). Support for T.L. from the Dutch Astrochemistry Network financed by The Netherlands Organisation for Scientific Research (NWO) is gratefully acknowledged.